\documentclass[]{pasj01}

\usepackage{url}

\Received{}
\Accepted{}
 
\begin{document} 

\title{ 
The activity of the post-nova V1363\,Cyg on long timescales }


\author{Vojt\v{e}ch \textsc{\v{S}imon}\altaffilmark{1,2}%
\thanks{This research made use of the AAVSO database.}}
\altaffiltext{1}{Astronomical Institute of the Czech Academy of Sciences, 25165 Ond\v{r}ejov, Czech Republic}
\altaffiltext{2}{Czech Technical University in Prague, Faculty of Electrical Engineering, 16627 Prague, Czech Republic}
\email{simon@asu.cas.cz}






\maketitle

\begin{abstract}
V1363\,Cyg is  a cataclysmic  variable  (CV) and a post-nova. Our analysis of 
its long-term optical activity used the archival data from the AAVSO database 
and literature.  We showed  that the accretion disk of V1363\,Cyg is  exposed 
to the thermal-viscous instability (TVI)  for at  least part of the time. The 
time fraction spent in the high state or the outbursts  dramatically  changed 
on the  timescale  of  decades.  Highly  variable  brightness  of  V1363\,Cyg 
displayed several episodes of a strong brightening (bumps in the light curve) 
from  a cool  disk  in  the  TVI  zone. In the  interpretation, their  vastly 
discrepant decay rates show that only some of these  bumps can  be attributed 
to the dwarf nova outbursts without strong irradiation of the disk by the hot 
white dwarf. The Bailey  relation  of  the  decay  rate, if  ascribed to a DN 
outburst of V1363\,Cyg, speaks in favor of its orbital period  $P_{\rm  orb}$ 
very long for a CV, about 20--40\,hr. A dominant cycle length of about 435\,d 
was present in the brightness changes all the time, even when  the  disk  was 
well inside the  TVI zone. We interpret it  as modulation  of the companion's 
mass outflow by differential rotation of the active region(s). 
\end{abstract}

\section{Introduction}

     Cataclysmic variables  (CVs) are  close  binary  systems in which matter 
transfers onto the white dwarf (WD) from its companion, a  lobe-filling  star 
(the secondary, the  donor)  (e.g.,  \citet{War95}).  Their  orbital  periods 
$P_{\rm  orb}$  range from  minutes to  several  days  \citep{Rit03}. Various 
processes can be involved in the long-term activity of various CV types (e.g., 
\citet{War95}). 

     If the value of the time-averaged mass transfer rate  $\dot m_{\rm  tr}$ 
between the donor and the non-magnetized accretor of a given CV accreting via 
the accretion disk is  between  some  limits, the  disk  embedding  the WD is 
exposed to the thermal-viscous instability  (TVI)  \citep{Sma84,Can94,Ham98}. 
It leads to episodic  accretion  of  matter  from the  disk  onto the WD. The 
brightness of the optical, thermal emission of the disk  increases by several 
magnitudes in  such  outbursts, typically  lasting for  days  or  weeks. Such 
CVs  are  called  dwarf  novae  (DNe)  (e.g., \citet{War95}). The TVI and the 
variations  of  $\dot  m_{\rm  tr}$ between the components play a significant 
role in CVs' activity. 

     CVs called novalikes often possess the high states in which no outbursts 
occur. Their absolute magnitudes $M_{\rm  opt}$ are high, comparable to those 
of the DN outbursts or even higher  \citep{War87}. Their  accretion disks can 
be ionized in these hot states \citep{War95}. The variations of $\dot  m_{\rm 
tr}$ govern the changes of $M_{\rm opt}$. 

     The donor's  magnetic  activity  can also  influence  the accretion disk 
structure \citep{Pea97}. It removes angular momentum  from the disk material, 
increasing the inward mass flow. 

     The magnetic  field  of  the  WD  strongly  influences the mass flow and 
activity of some CVs. Polars are CVs with  so strongly  magnetized  WD  ($B > 
10^{7}$\,Gauss) that the  accretion  disk cannot  form. Therefore, the matter 
flows directly  to  the  caps  at  the magnetic  poles  of  the  WD.  Various 
processes  (cyclotron,  bremsstrahlung)  produce  emissions of  the accretion 
column  of  the  matter  accreting  onto  the  WD (e.g.,  \citet{War95}). The 
high and low states typically last for weeks to months (e.g., \citet{War95}). 

     An explosion of a classical  nova is caused by episodic hydrogen burning 
of the accreted matter on the  WD \citep{Bod89,War95}. The source of nova can 
be either a CV with the accretion disk or a polar. 

    Extensive changes in the activity after the classical nova (CN) outbursts 
are observed in some CVs \citep{Liv87}. The features  consistent  with the DN 
outbursts (the TVI already operating) appear at least  intermittently, mainly 
in the later phases  (years, decades)  of the slow decay of the luminosity of 
some post-novae (e.g., V446\,Her \citep{Sti71,Hon11}). It suggests  that they 
contain the  accretion disks  able to  switch from the hot (ionized) state to 
the TVI regime on the timescale of decades.

     V1363\,Cyg (VV\,279), classified as a DN in the catalog of \citet{Bru87}, 
is a post-nova because it is approximately  centered  on a  nebula with about 
2\,arcmin diameter, ascribed to the ejecta of a  nova outburst \citep{Sah15}. 
The typical  duration  of  the  detectability  of  a nova  shell  is  several 
centuries  \citep{Sah15}, but  it  depends  on  several factors, e.g., on the 
distance of the nova. The  distance of  V1363\,Cyg $d = 1693 \pm 146$\,pc was 
determined from the observations by the satellite {\it Gaia} \citep{Gai18,Bai18} 
\footnote{\url{http://vizier.cfa.harvard.edu/viz-bin/VizieR?-source=I/347}}. 

     The photographic light  curve  of  V1363\,Cyg  in  1948--1955 showed the 
variations  probably  caused  by  the  DN  outbursts  and  state  transitions 
\citep{Mil71}.  \citet{Szk14} detected quasi-periodic oscillations during the 
decline  from  a  brightening  (outburst)  of  V1363\,Cyg.  The  spectrum  of 
V1363\,Cyg \citep{Bru92}  showed  strong  Balmer  emission lines  with  small 
decrement. They  also  identified  a Na\,D  absorption,  which indicates  the 
contribution  of the secondary  star both in the  continuum and line emission. 
The  continuum  was  inclined  to  the blue  at  short  and  the  red at long 
wavelengths. 

    In this paper, we  investigate  the  evolution  of the  long-term optical 
activity of V1363\,Cyg. A preliminary version of this analysis  was presented 
by \citet{Sim21}.

\begin{figure}
 \begin{center}
 \includegraphics[width=0.49\textwidth]{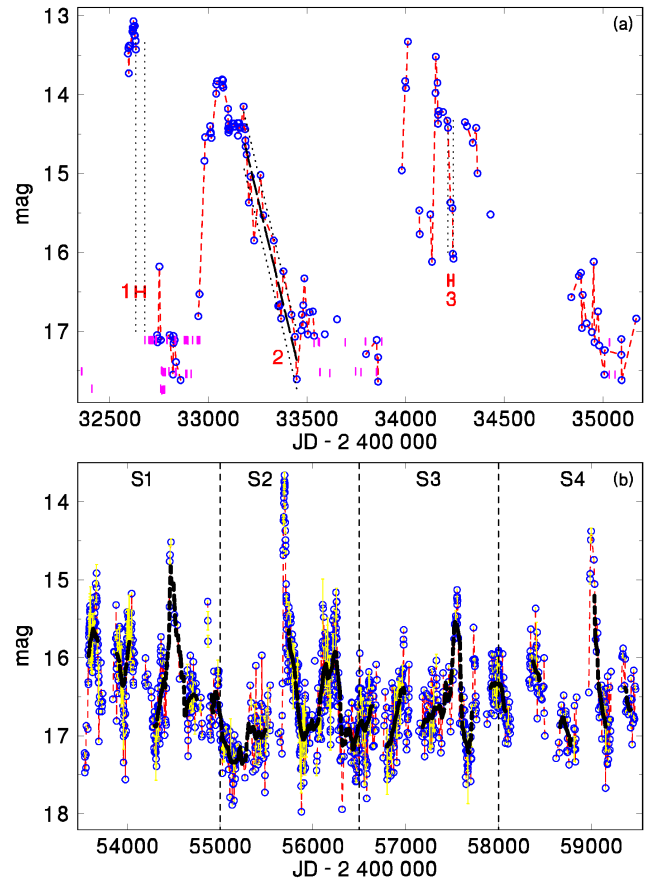}
 \end{center}
\caption{The long-term light curves of V1363\,Cyg. {\bf (a)} 
Set\,A. Number 2 denotes the fit to the decaying branch of a 
long brightening and its standard deviation. Numbers 1 and 3 
denote the limits of the steep decays. {\bf (b)} Set\,B. The 
error bars  of  the 1-d means are marked, but they are often 
smaller  than  the  size  of  the  symbol. The  points  were 
connected by  a line  in the densely  covered  parts  of the 
light curve to guide the  eye. The black line  marks the MAs 
for  $Q = 40$\,d. The  light  curve  was divided  into  four 
segments  (S1  to  S4).  The  distances  between  the  ticks 
on the  vertical  axes are the same  for  both diagrams. See 
Sect.\,\ref{ana} for  details.  (This figure is available in 
color in electronic form.) }
\label{1363all}
\end{figure}

\section{Observations      \label{obs}   }

                   The  AAVSO  International  database  (Massachusetts,  USA) 
\footnote{\url{https://www.aavso.org/data-download}}      \citep{Kaf19,Kaf21} 
contains both the  visual  and CCD  measurements  of V1363\,Cyg. We  used the 
$V$-band and unfiltered (with a sensitivity similar to the $V$-band) CCD data. 
The data obtained in these filters represented most of the  CCD  observations 
of V1363\,Cyg.  We included  visual  measurements  to  complete the  coverage 
of the  profile  of  a specific  feature. They  agreed  with  CCD  data.  The 
photographic data of  V1363\,Cyg  published  by \citet{Mil71}  enabled  us to 
extend the coverage of the light curve to the past.

\section{Data analysis    \label{ana}   }

\begin{figure}
 \begin{center}
 \includegraphics[width=0.49\textwidth]{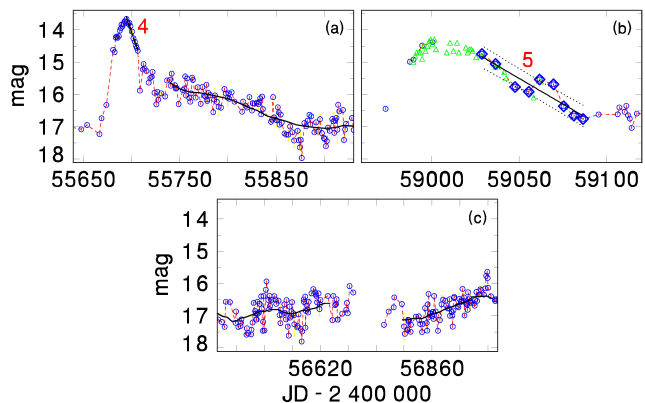}
 \end{center}
\caption{{\bf  (a)}  A bright outburst  with a very long  decay in 
V1363\,Cyg. Circles represent  the  1-d means  of  the  AAVSO  CCD 
$V$-band and unfiltered data. Also, the  MAs of the  outburst tail 
are shown.  {\bf  (b)}  A bright, well-defined  outburst. The  1-d 
means of the  AAVSO  visual measurements (triangles)  are included 
to cover the outburst  profile. {\bf  (c)}  Example of the segment 
and the MAs of the light curve without brightenings. The distances 
between the  ticks  on  the  vertical  axes  are  the same for all 
diagrams. The straight lines with  the standard deviation show the 
fits to the  decaying  branches  (abbreviated  as  4  and  5). See 
Sect.\,\ref{ana} for details. (This figure  is  available in color 
in electronic form.)         }
\label{segm}
\end{figure}

\subsection{Overview of the activity  \label{ove} }  

     The long-term activity of V1363\,Cyg was  revealed  to consist mainly of 
a series of  large-amplitude  fluctuations  (set\,A  (the  1-d  means  of the 
photographic data of \citet{Mil71}) in  Fig.\,\ref{1363all}a  and set\,B (the 
1-d means of the $V$-band and unfiltered AAVSO data) in Fig.\,\ref{1363all}b). 
An  inspection  showed  a typical  uncertainty  of  a 1-d  mean of set\,A was 
0.1--0.2\,mag. Regarding the  AAVSO  data in  Fig.\,\ref{1363all}b,  only the 
well-covered part  of the  light  curve, abbreviated as set\,B, was selected. 
The coverage by the detections was dense, with only the  short seasonal gaps. 
The dominant brightness variations occurred  on  the  timescale  of days  and 
weeks. This segment is about twice longer than set\,A in Fig.\,\ref{1363all}a.

\begin{figure}
 \begin{center}
 \includegraphics[width=0.48\textwidth]{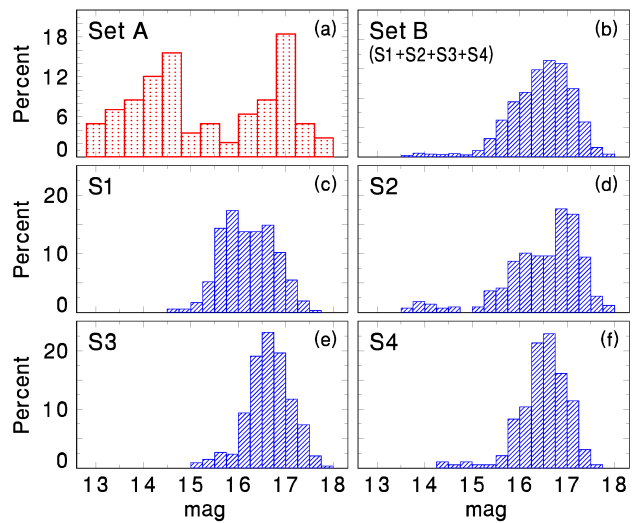}
 \end{center}
\caption{Histograms of the brightness of V1363\,Cyg in various time 
segments. {\bf (a)} The photographic data from Fig.\,\ref{1363all}a 
(set\,A). {\bf (b)}  Set\,B  (segments S1 to  S4) of the AAVSO data 
from Fig.\,\ref{1363all}b. {\bf  (c, d, e, f)} Segments S1--S4. See 
Sect.\,\ref{ana} for details. (This figure is available in color in 
electronic form.) }
\label{hist}
\end{figure}

     Usually, several  CCD  images  of  V1363\,Cyg  were obtained  per night. 
The long series of more than 200\,images were  very few. We assessed how much 
such brightness changes affected the light curve on the timescale of days and 
longer. Therefore, we calculated  the brightness  means  from the data points 
obtained in a given night. This approach enabled us to calculate the standard 
deviation of the  brightness of such  a night mean. We found that the scatter 
of the individual observations  obtained in  a given night  was  smaller than 
that occurring on the timescales of days and weeks (Fig.\,\ref{1363all}). The 
night means thus provided  a good  representation  of the light curve on long 
timescales. 

     Most CCD observations of V1363\,Cyg were obtained in the $V$-band, while 
Miller's (1971) photographic data band could be approximated by the $B$-band. 
Although we compared the $B$-band light curve with that of the  $V$-band, the 
peak-to-peak amplitude of the brightness variations  was  considerably higher 
than the possible  changes of the  color  indices. The color  variations were 
thus not expected to influence our results dramatically. 

     The two-sided moving  averages  (MA)  of the brightness of CCD data were 
made to separate  the long-term  brightness  variations  from  the superposed 
rapid  (e.g., night-to-night)  fluctuations. This  method  was  described  by 
\citet{Bro87}. Various values of  the filter  half-width  $Q$  (in days) were 
tested. Figure\,\ref{1363all}b displays the light curve for $Q = 40$\,d, step 
10\,d. The MAs were interrupted in the seasonal gaps and their vicinity. Also, 
the sharp peaks  of some  strong brightenings (Fig.\,\ref{segm}) were omitted 
from this fitting  because  $Q = 40$\,d  yielded a good fit to the relatively 
gradual variations but not to the occasional sharp peaks. 

     Figure\,\ref{segm} shows  the  examples  of the  features  of the  light 
curve of V1363\,Cyg. The  very  bright  outbursts  in  Figs.\,\ref{segm}a and 
\ref{segm}b differ  from each  other by the profile of the decaying branch (a 
very long  one  with  strong  undulations is displayed in Fig.\,\ref{segm}a). 
Figures\,\ref{segm}c and  \ref{segm}d  show the examples of  the  complicated 
light  curve of V1363\,Cyg in the time segments without intense brightenings. 
The ``quiescent" levels (plateaux) can attain various brightnesses. 

     The  histograms  of  the  brightness  of   V1363\,Cyg  in  various  time 
segments are displayed in  Fig.\,\ref{hist}.  The  histogram  for set\,A from 
Fig.\,\ref{1363all}a was compared with that for set\,B from Fig.\,\ref{1363all}b. 
The upper limits in the photographic data could only influence the brightness 
lower than 17\,mag. Further, the whole set\,B was divided into four  segments 
S1--S4 to assess the time evolution of the brightness in Figs.\,\ref{hist}cdef. 

     A comparison  with  histogram  for  brightnesses  of  set\,A shows a big 
difference in  a  part  of  the   brightness  higher  than  15\,mag. In  this 
brightness range, Fig.\,\ref{hist}a contains a big bump where Fig.\,\ref{hist}b 
displays only a long tail. 

     The histogram for the brightnesses of set\,B usually consists of a broad, 
asymmetric bump. The  prominence  of the  tail  toward  the higher magnitudes 
varied for the individual time segments. A very prominent tail was present in 
segment S2.

\begin{figure}
 \begin{center}
 \includegraphics[width=0.48\textwidth]{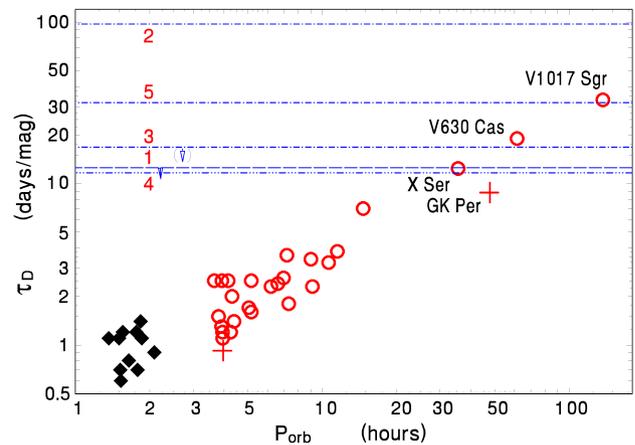}
 \end{center}
\caption{The decay rate of DN outbursts, $\tau_{\rm  D}$, 
versus $P_{\rm  orb}$. DNe  below  and  above  the period 
gap (e.g., \citet{War95})  are marked by the diamonds and 
circles,  respectively.  Crosses  mark  the  intermediate 
polars (see \citet{War95} for definition). The horizontal 
lines with the numbers  denote  the values  of $\tau_{\rm 
D}$ of the decays of events 1--5 in Figs.\,\ref{1363all}a 
and \ref{segm}. See Sect.\,\ref{ove} for details.  }
\label{ptd-xs}
\end{figure}

     The fits  to  the  decays  of  the  brightenings  numbered  as  1--5  in 
Figs.\,\ref{1363all}a and \ref{segm}, supplemented by the upper limits in the 
fragmentary data, enabled us to  determine their decay rates  $\tau_{\rm  D}$ 
(in d\,mag$^{-1}$). They served the analysis  of the nature of these  events, 
discussed in Sect.\,\ref{dis}. A comparison of the decaying branches of these 
brightenings with those of DNe is shown in Fig.\,\ref{ptd-xs}; $\tau_{\rm D}$ 
is plotted versus  $P_{\rm  orb}$. It enables us to discuss the properties of 
the episodic brightenings (outbursts). Moreover, an assessment of whether the 
Bailey  relation of DN  outbursts \citep{Bai75} applies to any of them can be 
made.  Although  the  coverage  of  some  events is too fragmentary to obtain 
reliable values of  $\tau_{\rm  D}$ of their decays, even the upper limits of 
the  decay  rate  enable  us  to show  the broad range of  $\tau_{\rm  D}$ in 
V1363\,Cyg. 

     The data in  Fig.\,\ref{ptd-xs} come from  \citet{War95}, \citet{Sim00a} 
(DO\,Dra), \citet{Sim00b} (DX\,And), \citet{She10} (V630\,Cas), \citet{Sim18} 
(X\,Ser), AAVSO database \citep{Kaf21} (V392\,Per, UY\,Pup). Please note that 
the values of $\tau_{\rm  D}$ of DO\,Dra and GK\,Per are smaller than in most 
DNe with the comparable $P_{\rm orb}$. For GK\,Per, only the steepest part of 
the decline was used for the determination of $\tau_{\rm  D}$  \citep{Sim18}. 
These two CVs are known to be the intermediate polars, according to \citet{Pat92} 
and \citet{Wat85}. 

   The horizontal lines in Fig.\,\ref{ptd-xs} denote the values of $\tau_{\rm 
D}$ of events 1--5 of  V1363\,Cyg from Figs.\,\ref{1363all}a  and \ref{segm}. 
The belt of  $\tau_{\rm  D}$  of DNe  (diamonds,  circles, crosses), spanning 
between $P_{\rm orb}$ of 1.1\,hr and 137\,hr, indicates the length of $P_{\rm 
orb}$ of V1363\,Cyg bigger than about 20\,hr if its brightening episodes with 
the smallest  $\tau_{\rm  D}$  (events  1, 3, and  4) are  the  DN  outbursts 
obeying the Bailey relation \citep{Bai75}.

\subsection{Time variations     \label{time} }  

\begin{figure}
 \begin{center}
 \includegraphics[width=0.48\textwidth]{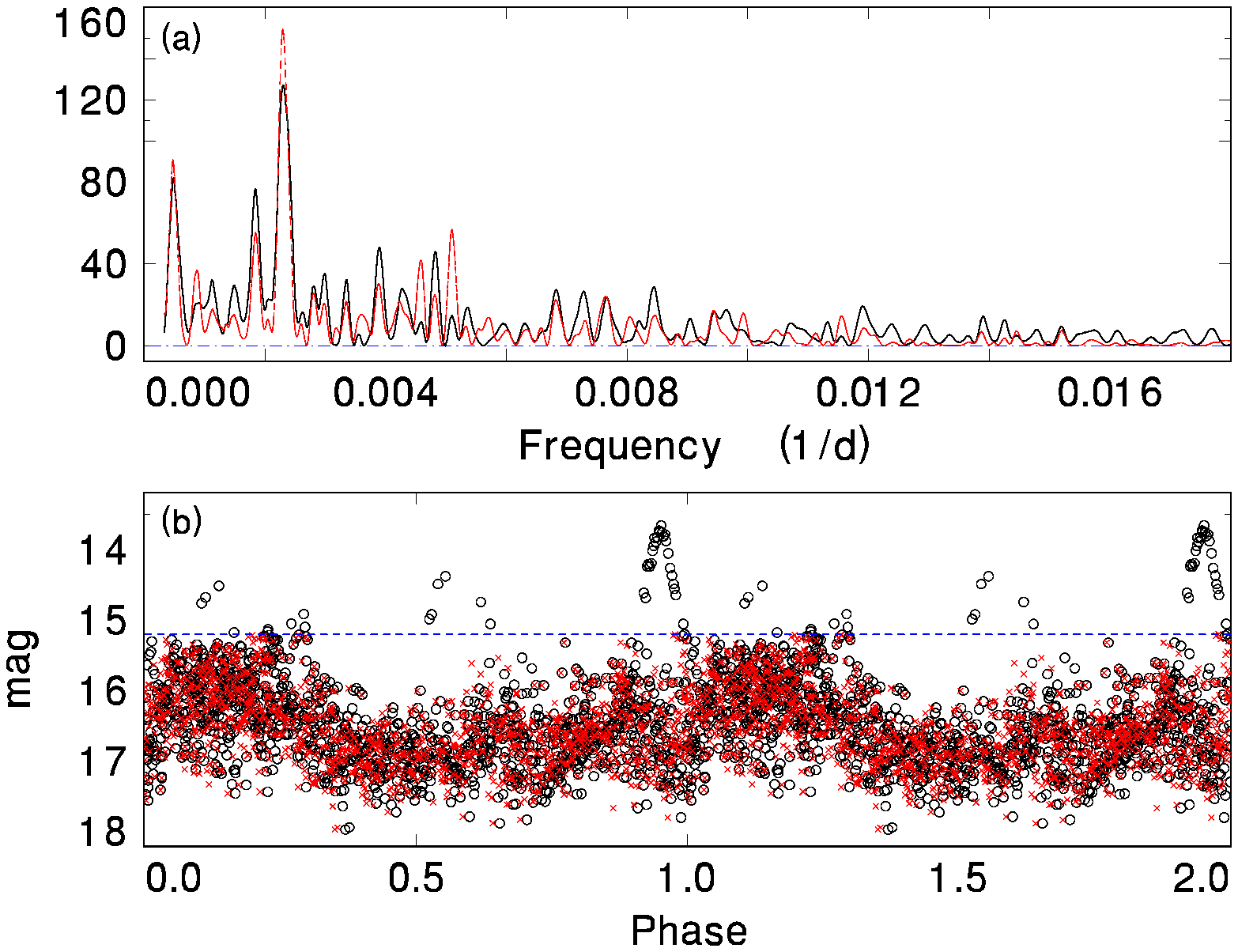}
 \end{center}
\caption{Periods   $CL$  of   the  brightness  variations  in 
V1363\,Cyg. The data were investigated with the  Lomb-Scargle 
method \citep{Lom76}. {\bf (a)} Periodogram for the 1-d means 
of the brightness (a solid line). The red dashed line denotes 
a periodogram  for  the  data  with  brightness   lower  than 
15.2\,mag to avoid flares. {\bf  (b)}  The 1-d  means  of the 
brightness folded with  $CL = 434.8$\,d  (the highest peak of 
the solid line in panel {\bf  (a)}). The red {\sf  x} denotes 
the data with brightness lower than 15.2\,mag folded with $CL 
= 436.3$\,d (the highest peak of the red dashed line in panel 
{\bf (a)}). The starting point of the folding is the start of 
both data sets in JD\,2\,453\,541.8 in  Fig.\,\ref{1363all}b. 
The folded light curve is plotted twice. See Sect.\,\ref{ana} 
for details. (This figure is available in color in electronic 
form.)    }
\label{per}
\end{figure}

    A search for periods in the brightness variations in Fig.\,\ref{1363all}b 
used the Lomb-Scargle method \citep{Lom76}, included in the code Peranso\footnote{www.peranso.com}. 
The  1-d  means  of  the  brightness  from  Fig.\,\ref{1363all}b)  were  used 
(Fig.\,\ref{per}a). 

     We found the modulation  with the  period (cycle-length)  $CL$  of about 
435\,d. The folded  light curve  of  the  1-d  means  in  Fig.\,\ref{per}b is 
gradual, roughly symmetric. For comparison, a period search of the  data with 
brightnesses  lower  than  15.2\,mag to avoid  flares was made. The resulting 
periodograms are included in Fig.\,\ref{per}a. Both  the periodograms and the 
folded light curves are mutually similar for both sets (Fig.\,\ref{per}).

\begin{figure}
 \begin{center}
 \includegraphics[width=0.48\textwidth]{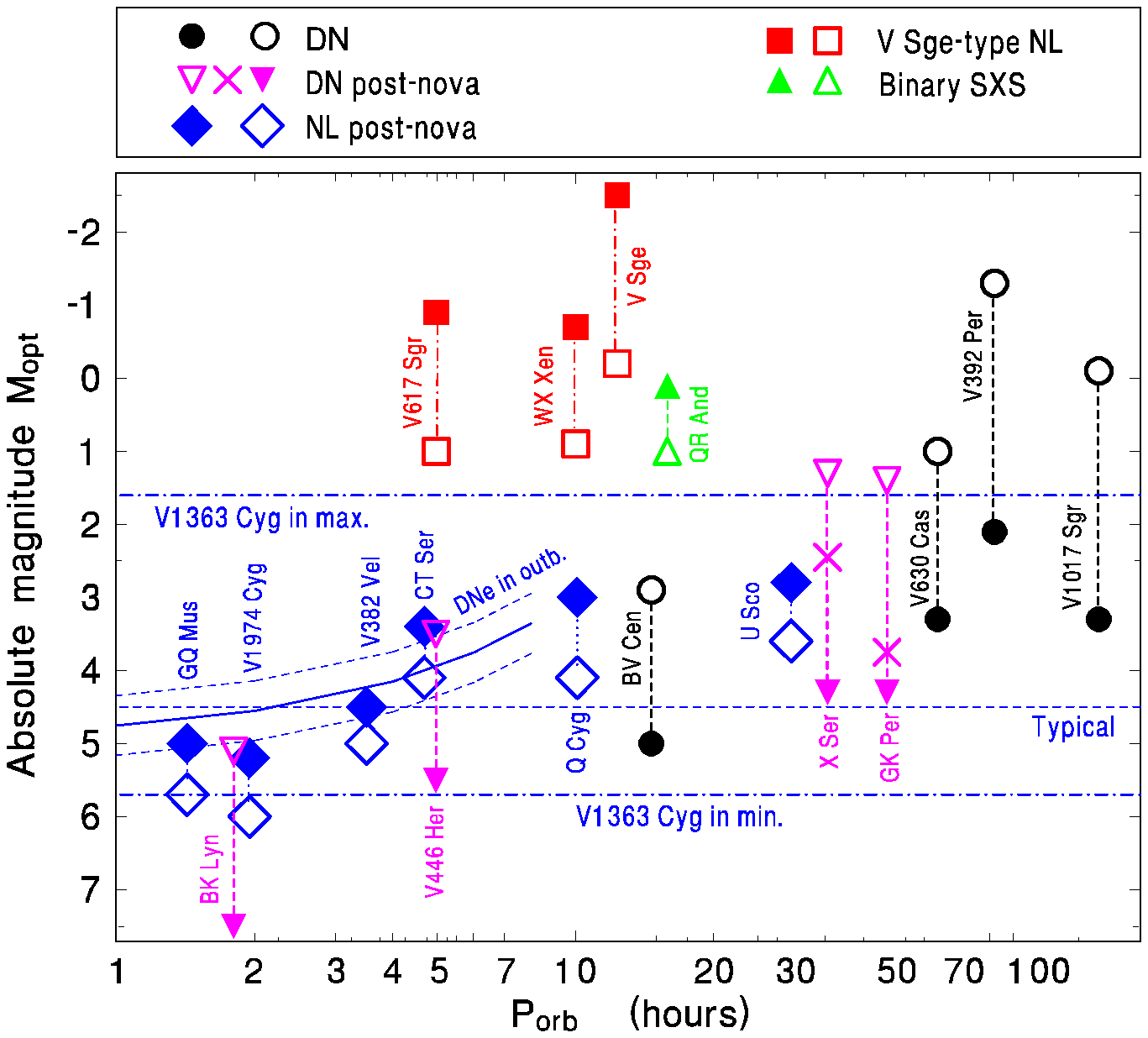}
 \end{center}
\caption{The  positions  of  V1363\,Cyg (magnitudes in  CCD data) in 
the $M_{\rm  opt}$ versus  $P_{\rm  orb}$  diagram. Its  maximum and 
minimum values of $M_{\rm opt}$ are given. The typical $M_{\rm opt}$ 
of V1363\,Cyg corresponds to the  peak  of the bump in the histogram 
in Fig.\,\ref{hist}b. The maximum and the minimum magnitudes (except 
deep eclipses  in  U\,Sco  \citep{Sch95})  of CVs are displayed. The 
solid blue line represents maxima's and  supermaxima's  peaks of DNe 
with $P_{\rm orb} \leq 8$\,hr \citep{Pat11}. The dashed lines denote 
the standard deviation. The meanings of the symbols are given in the 
legend (they  include  the  observed  outbursts  of DNe with $P_{\rm 
orb}$  longer  than  those in  \citet{Pat11}.  See  Sect.\,\ref{ana} 
and Table\,\ref{tab:1} for  details.  (This  figure  is available in 
color in electronic form.)    }
\label{abs}
\end{figure}

\begin{table*}
      \tbl{Parameters and types of CVs used in  Fig.\,\ref{abs}. These objects are 
           arranged according to the length of $P_{\rm  orb}$. N is  nova, RN is a 
           recurrent nova, NL is novalike, VS is V\,Sge-type, DN is dwarf nova, UG 
           is DN of the U\,Gem  type, UGZ is DN  of the  Z\,Cam-type. LC refers to 
           the sources of  the light  curves  for  Fig.\,\ref{abs}. The lengths of 
           $P_{\rm  orb}$ are  measured  in  hours. The  highest  and  the  lowest 
           apparent magnitudes  are abbreviated  as $mag_{\rm  max}$ and $mag_{\rm 
           min}$. They refer to the  pre-nova  or post-nova. The  highest  and the 
           lowest absolute magnitudes are abbreviated as $M_{\rm max}$ and $M_{\rm 
           min}$.  Distance,  denoted  as  $d$,  is  given in  parsecs. Extinction 
           measured in magnitudes is abbreviated as $A_{V}$. See Sect.\,\ref{cont} 
           for details. 
  \footnotemark[$*$] }{%
  \begin{tabular}{llllllllll}
      \hline
      Name       &           Type             & LC    & $P_{\rm orb}$ (hr) & $mag_{\rm max}$ & $M_{\rm max}$ &  $mag_{\rm min}$ &  $M_{\rm min}$ & $d$                & $A_{V}$  \\ 
      \hline
GQ\,Mus    & N(LO83),NL(AAVSO)          & AAVSO &     1.425 DS89     &      18.2       &      5.0      &      18.9        &      5.7       & 2456    & 1.2 K84   \\
BK\,Lyn    & N(H58), NL, DN (P13)       &  P13  &     1.7995 R96     &      13.6       &      5.1      &      16.0        &      7.5       &  498    & 0   \\
V1974\,Cyg & N(C92),NL?(SC10)           & AAVSO &     1.9503 D94     &      16.8       &      5.2      &      17.6        &      6.0       & 1568    & 0.6       \\
V382\,Vel  & N(G00),NL(AAVSO)           & AAVSO &     3.5076 B01     &      16.0       &      4.5      &      16.5        &      5.0       & 1707    & 0.3  D02  \\
CT\,Ser    & N(M48),NL(R05,CRTS)        & CRTS  &     4.68 R05       &      16.1       &      3.4      &      16.8        &      4.1       & 3453    & 0.01      \\
V446\,Her  & N, NL(S71), UG(H11)        & H11   &     4.97 TT00      &      15.1       &      3.5      &      17.1        &      5.5       & 1308    & 1.0       \\
V617\,Sgr  & VS(SD98)                   & AAVSO &     4.9718 C99     &      14.5       &     -0.9      &      16.4        &      1.0       & 9539    & 0.5       \\
WX\,Cen    & VS(SD98)                   & AAVSO &    10.0 DS95       &      12.7       &     -0.7      &      14.3        &      0.9       & 2699    & 1.2 SD98  \\
Q\,Cyg     & N, NL(K03)                 & AAVSO &    10.08 K03       &      14.2       &      3.0      &      15.3        &      4.1       & 1320    & 0.6 KS11  \\
V\,Sge     & NL(H65),VS(SD98)           & AAVSO &    12.3406 H65     &       9.9       &     -2.5      &      12.2        &     -0.2       & 2236    & 0.7       \\
BV\,Cen    & DN(B74)                    & AAVSO &    14.631 VB80     &      11.1       &      2.9      &      13.2        &      5.0       & 363     & 0.4 W880  \\
QR\,And    & SX,NL(GW95)                & GW95  &    15.85 B95       &      11.6       &      0.1      &      12.5        &      1.0       & 1857    & 0.2       \\  
U\,Sco     & RN(W81)                    & AAVSO &    29.5335 SR95    &      17.6       &      2.8      &      18.4        &      3.6       & 6559    & 0.74 P15  \\
X\,Ser     & N(L08,D81,D90),NL(H98),UG(S18) & S18   &    35.472 TT00     &      14.2       &      1.3      &      17.2        &      4.3       & 3056    & 0.5       \\ 
GK\,Per    & N(SB83),NL,UG(SB83,H81)    & S02   &    47.76 C86       &      10.3       &      1.4      &      13.2        &      4.3       &  437    & 0.7       \\  
V630\,Cas  & UG(S17,SP10)               & SP10  &    61.5329 O01     &      13.9       &      1.0      &      16.2        &      3.3       & 3004    & 0.5       \\ 
V392\,Per  & N,NL(M20),UG(AAVSO)        & AAVSO &    81.8832 M20     &      13.5       &     -1.3      &      16.9        &      2.1       & 3416    & 2.1       \\  
V1017\,Sgr & UG,N(M46)                  & M46   &   138.144 S17      &      10.9       &     -0.1      &      14.3        &      3.3       & 1227    & 0.6       \\
V1363\,Cyg & UGZ(M71,AAVSO)             & AAVSO &                    &      13.7       &      1.6      &      17.8        &      5.7       & 1693    & 1.0       \\
      \hline
    \end{tabular}}\label{tab:1}
\begin{tabnote}
\footnotemark[$*$] References: 
LO83: \citet{Lil83}; 
AAVSO \citep{Kaf21}; 
DS89: \citet{Dia89}; 
K84: \citet{Kra84}); 
  H58: \citet{Hsi58}; 
  P13: \citet{Pat13}; 
  R96: \citet{Rin96}; %
C92: \citet{Col92}; 
SC10: \citet{Sch10}; 
D94: \citet{DeY94}; 
G00: \citet{Gar00}; 
B01: \citet{Bos01}; 
D02: \citet{Del02}; 
M48: \citet{McL48}; 
R05: \citet{Rin05}; 
CRTS: \citet{Dra09}; 
S71: \citet{Sti71}; 
H11: \citet{Hon11}; %
SD98: \citet{Ste98}; 
C99: \citet{Cie99}; 
DS95: \citet{Dia95}; 
K03: \citet{Kaf03}; 
KS11: \citet{Kol11}; 
H65: \citet{Her65}; 
B74: \citet{Bat74}; 
W880: \citet{Wil88}; 
GW95: \citet{Gre95}; 
B95: \citet{Beu95}; 
W81: \citet{Wil81}; 
SR95: \citet{Sch95}; 
P15: \citet{Pag15}; 
L08: \citet{Lea08}; 
D81: \citet{Due81}; 
D90: \citet{Due90}; 
H98: \citet{Hon98}; 
S18: \citet{Sim18}; 
TT00: \citet{Tho00}; 
SB83: \citet{Sab83}; 
H81: \citet{Hud81}; 
S02; \citet{Sim02}; 
C86: \citet{Cra86}; 
S17: \citet{Sam17}; 
SP10: \citet{She10}; 
O01: \citet{Oro01}; 
M20: \citet{Mun20}; 
M46: \citet{McL46}; 
S17: \citet{Sal17}; 
M71: \citet{Mil71}; 
  \\ 
\end{tabnote}
\end{table*}

\subsection{V1363\,Cyg in the context of the CV absolute magnitudes   \label{cont} }

     Figure\,\ref{abs}  shows  the  position of  V1363\,Cyg  in  the absolute 
optical  magnitude  $M_{\rm  opt}$ vs. $P_{\rm  orb}$  diagram  of  CVs  with 
accretion disks. The range of $M_{\rm  opt}$ similar to that of V1363\,Cyg in 
various stages of its  activity is plotted. Table\,\ref{tab:1} summarizes the 
parameters of  these  CVs. The values  of their  $d$ were determined from the 
observations by ESA {\it Gaia} \citep{Bai18} 
\footnote{\url{http://vizier.cfa.harvard.edu/viz-bin/VizieR?-source=I/347}}. 
Extinction $A_{V}$  mainly  was  determined  from  the  3D  map  of Galactic 
reddening  \footnote{\url{http://argonaut.skymaps.info/}}  of \citet{Gre18}. 

   The zone below maxima's peaks of outbursts of DNe \citep{Pat11} represents 
the region in which the  accretion  disks  are  exposed to the TVI. Also, the 
minimum and maximum magnitudes  of several  DNe (as the  AAVSO  light  curves 
show) with $P_{\rm  orb}$ longer than those in \citet{Pat11} are displayed in 
Fig.\,\ref{abs} for comparison. These peak magnitudes enable us to extend the 
TVI region to $P_{\rm  orb}$  longer than used by \citet{Pat11}. An extension 
of his fit from  the 6--8\,hr region to the longer  $P_{\rm  orb}$ shows that 
all  CVs in  Fig.\,\ref{abs}  lying  below it are in the TVI region. Only the 
outburst peak of V392\,Per may approach its limit. 

     Figure\,\ref{abs}  also shows post-novae of various types  with  $M_{\rm 
opt}$ in the zone of brightnesses similar to that of V1363\,Cyg. The {\sf  x} 
symbols in Fig.\,\ref{abs} denote the states of  activity  of  the post-novae 
X\,Ser \citep{Hon98} and GK\,Per \citep{Kaf21}, in  which the quiescent level 
was brighter than usual, and the DN outbursts  were suppressed or absent. The 
recurrent nova U\,Sco inside the TVI region displays no DN outbursts \citep{Kaf21}. 
The DN outbursts of V392\,Per and V1017\,Sgr were observed only  before their 
CN outbursts (AAVSO; \citet{McL46}). V1363\,Cyg itself is situated in the TVI 
region in quiescence (near min. brightness) irrespective of its $P_{\rm orb}$.

\section{Discussion    \label{dis}     }

    We present an analysis of the long-term optical activity of the post-nova 
V1363\,Cyg. We show  that  its  disk is  exposed  to the  TVI at  least  near 
the minimum  brightness. The  time  fraction  spent in  the high state or the 
outbursts  dramatically changed  on  the  timescale of decades, as shown by a 
comparison of  the  histogram  for  the  brightness  of set\,A with those for 
set\,B (Fig.\,\ref{hist}). 

    We interpret the histogram of the brightness in set\,B (Fig.\,\ref{hist}), 
consisting of a broad bump with a long tail toward  the higher brightness, as 
the fluctuations and occasional brightenings from a cool disk in the TVI zone. 
The value of $M_{\rm opt}$ in the outburst peak of V1363\,Cyg speaks in favor 
of $P_{\rm  orb} > 12$\,hr, as indicated  by the  peak  magnitudes  of the DN 
outbursts  in  \citet{Pat11}  (Fig.\,\ref{abs}).

     The detectability of the secondary as the Na\,D absorption \citep{Bru92} 
speaks in favor of  $P_{\rm  orb}$ of V1363\,Cyg considerably longer than the 
period gap  (e.g., \citet{Dan82}). A contribution of the  secondary component 
to the light curve profile may be significant, especially near the lower part 
of the  histograms in  Fig.\,\ref{hist}. Since  the  brightness of V1363\,Cyg 
only seldom  achieves its  lowest  level, V1363\,Cyg appears to be active all 
the time. The value $M_{\rm min} \approx 5.7$ (Table\,\ref{tab:1}) shows that 
if the fluxes of the secondary  and the disk are  mutually comparable in this 
epoch, the  position  of  V1363\,Cyg  in Fig.\,\ref{abs}  may  be  similar to 
that of BV\,Cen  with  a G5-G8\,IV-V secondary  \citep{Vog80}. As regards the 
luminosity of this component, the current  type of  CV is  determined  by its 
previous evolution \citep{Pod03}. The  spectra  of V1363\,Cyg,  especially in 
its low brightness seasons, can help. 

     A dominant cycle-length of about 435\,d, much longer than the superposed 
features like outbursts, was present in the light curve of V1363\,Cyg all the 
time in set\,B, even when the  disk  was  well  inside  the  TVI  region. The 
outbursts  or  flares  occurred irrespective  of  the  phase  of  this  cycle 
(Fig.\,\ref{per}). In the interpretation, it can be caused  by the appearance 
and changes of  the position  of  the  active  regions  (loops \citep{Kaf08}, 
starspots  \citep{Liv94})  with  respect  to the  L1 point  by a differential 
rotation of the lobe-filling donor  \citep{Sch82}. It  can cyclically  modify 
the donor's mass outflow  and the mass transfer to the disk. The brightenings 
caused by the TVI or the mass transfer bursts were superimposed on this cycle 
length of about 435\,d. 

     We  found  vastly  discrepant  values  of  $\tau_{\rm  D}$  for  various 
brightening  episodes  of  V1363\,Cyg.  The  quantity  $\tau_{\rm  D}$  is an 
essential parameter that helps resolve various processes governing the event. 
We interpret bumps No.\,1, 3, and 4  with  the smallest  and mutually similar 
values (or  the upper  limits)  of  $\tau_{\rm  D}$  (Fig.\,\ref{ptd-xs})  as 
a propagation of  the cooling front  in  the  disk  finishing the DN outburst  \citep{Sma84,Ham98}. In this  context, event\,4 (Fig.\,\ref{segm}a) is  not a 
simple DN outburst. We interpret its initial decay  phase as a propagation of 
the cooling front  similar  to events\,1 and 3. A significant  flattening and 
undulations in the  more  advanced part of the decay can be  explained  by an 
increase of the mass outflow from the donor's nozzle  due to its  irradiation 
in the advanced outburst phase (see the model of \citet{Ham00}). 

   A comparison of the mutually similar values of $\tau_{\rm D}$ of events\,1, 
3, and 4 with the position of V1363\,Cyg in the $\tau_{\rm D}$--$P_{\rm orb}$ 
diagram  for  DNe  (Fig.\,\ref{ptd-xs})  indicates its  $P_{\rm  orb}$  about 
20--40\,hr, provided that the TVI causes them and  obeys the  Bailey relation 
\citep{Bai75}. Also, the  peak  magnitude  of  V1363\,Cyg  (the  tail  of the 
histogram  in Fig.\,\ref{hist}b) shows that  this  CV would be located in the 
TVI region all the time if its  $P_{\rm  orb}$  were longer than about 20\,hr 
(Fig.\,\ref{abs}). 

     On the contrary, the divergent and remarkably high values of  $\tau_{\rm 
D}$ of the broad events\,2 and 5 (discrete brightening  episodes still in the 
TVI region) need modification of the cooling front propagation  for these two 
cases. Event\,2 was roughly similar (although shorter) to a specific state of 
activity (several very broad bumps  (the  longest  one  600\,d)  in the light 
curve)  of  the  post-nova   X\,Ser  \citep{Hon98}  (the  {\sf  x}  symbol in 
Fig.\,\ref{abs}). Such  bumps  in  X\,Ser  \citep{Hon98}, occurring  among DN 
outbursts  \citep{Sim18}, had the peak  $M_{\rm  opt}$  still  inside the TVI 
region  (i.e., below  an  extrapolation  of  the  maxima's  peaks of DNe with 
$P_{\rm  orb}$ \citep{Pat11}) (Fig.\,\ref{abs}). A decrease of $M_{\rm  opt}$ 
of a post-nova  containing  the disk with time thus may not always be regular 
and irreversible. 

     A variation of $\dot m_{\rm tr}$ in a given DN does not lead to a change 
of $\tau_{\rm  D}$  of the DN  outbursts  \citep{Bua01}. However, irradiating 
the disk by the hot WD in a post-nova can modify the situation. \citet{Sch00} 
show that even if the accretion rate  in such a CV is  low  enough to  permit 
TVI, disk irradiation from  the very hot WD suppresses DN outbursts for about 
a century  since  the CN  outburst. Using  this  model, we  assume  that  the 
irradiated  disk remains in  the hot  state  during a significant part of the 
decaying  branch  of the  DN outburst. Its  optical  flux  will  decay as the 
disk's mass decreases  because  accretion  onto the WD will diminish with the 
outburst progress. The broad brightening episodes with slow  decays  can thus 
be caused  both  in the post-novae V1363\,Cyg and X\,Ser and keep the disk in 
the hot state for a big part of the decay of the DN outburst.

     In the interpretation, the remarkably low  values of  $\tau_{\rm  D}$ of 
events\,2 and 5 show that the structural changes of the  disk that can modify 
its irradiation  by  the  hot  WD  may  be  significant only  in part  of the 
outbursts in V1363\,Cyg. The changes of the disk influencing this irradiation 
can occur with various strengths all the time, as indicated  by the quiescent 
(out  of  the  outbursts)  brightness  variations  (Figs.\,\ref{1363all}  and 
\ref{per}). The  activity of  the active  regions on  the secondary component 
mentioned above may  contribute  to the modification of mass flow to the disk 
with time. The repeated observing of the variations of the orbital modulation 
in V1363\,Cyg can help assess the structural changes of  the  disk  and their 
variations. Such irradiated disk also deserves further modeling. 

     The absence of DN outbursts in most post-novae with $P_{\rm orb}< 8$\,hr 
in the zone of $M_{\rm opt}$ of V1363\,Cyg (Fig.\,\ref{abs}) shows that these 
systems reside on the  peak  magnitude of the outbursting  DNe  with the disk 
ionized out to its outer rim because of a high $\dot m_{\rm tr}$ \citep{Ham98}. 
Also, the position of the recurrent nova U\,Sco in Fig.\,\ref{abs} similar to 
that of GK\,Per between its quiescence and outburst peak is evidence that the 
post-novae  can  reside  without  the  DN outbursts  in the  TVI zone. In the 
context of the activity of V1363\,Cyg and other post-novae, irradiation  from 
the cooling WD  play  a significant role in  the post-nova  evolution and the 
character of activity \citep{Sch01}. This WD remains very hot ($T_{\rm  eff}$ 
a few $\times 10^{5}$\,K) after the nova explosion. This value only gradually 
decreases with time \citep{Pri86}. If the  disk is subject to the  TVI in the 
absence  of  irradiation,  strong  irradiation  by  the  WD  can suppress the 
TVI, while a weaker  one  gives  rise  to an  inner steady-state disk region, 
surrounded by a TVI unstable outer annulus \citep{Min90,Sch00,Sch01}. 

    In addition, the peak  $M_{\rm  opt}$  (including event\,2) of V1363\,Cyg 
was only about 1\,mag lower than those of the binary  supersoft  X-ray source 
(BSXS) QR\,And \citep{Beu95} and some V\,Sge-type  CVs  \citep{Ste98}. In the 
models  of \citet{Heu92,Tee98,Gin21}, irradiation of  the  companion  in BSXS 
can invoke a very high  $m_{\rm  tr}$. It may be at a self-sustaining rate of 
about  $10^{-7}$M$_{\odot}$\,yr$^{-1}$  for up  to a thousand years since the 
CN outburst in  some  cases  \citep{Gin21}. According  to \citet{Ste98}, some 
V\,Sge-type CVs may be the optical counterparts of BSXSs. 

     Extensive changes in the activity after the CN outbursts are observed in 
some CVs \citep{Liv87}. The  features consistent  with  the DN outbursts (the 
TVI already  operating) appear  at least  intermittently, mainly in the later 
phases  (years,  decades)  of the  slow  decay of the mean luminosity of some 
post-novae  (e.g.,  V446\,Her  \citep{Sti71,Hon11};  BK\,Lyn   \citep{Pat13}; 
GK\,Per  \citep{Hud81,Sab83}  X\,Ser  \citep{Sim18}).  It  suggests that they 
contain the accretion disks able to switch from  the hot  (ionized)  state to 
the TVI regime and produce the DN outbursts during the years or decades since 
the finish of  the CN  outburst. V1363\,Cyg  appears to be already in the TVI 
region, too. 

     There are also some other CVs that  have  observable shells suggesting a 
CN outburst in the system in the past: Z\,Cam, the limit of 1300\,years since 
the CN outburst  \citep{Sha12}, AT\,Cnc with the CN outburst about 330\,years 
ago {\citep{Sha17a},  Nova  Sco\,1437 \citep{Sha17b}. It constrains the upper 
limit of the time elapsed since the nova outburst also for V1363\,Cyg. The DN 
outbursts  show  that these  post-novae  are  already in the TVI zone. The CN 
explosion of V1363\,Cyg was thus a relatively recent event.

\begin{ack}
This study was supported by the EU project H2020 AHEAD2020, grant agreement 
871158. Also, support  by  the project  RVO:67985815 is  acknowledged. This 
research used the observations from the AAVSO  International database (USA).  
I acknowledge  with  thanks  the variable  star observations from the AAVSO 
International Database  contributed by observers worldwide and used in this 
research.
\end{ack}

\end{document}